# Characterization of $Tl_2NaYCl_6$:Ce Scintillation Crystals as Gamma-Ray Detectors

R. Hawrami[1], E. Ariesanti[2], A. Burger[2]
[1]*Xtallized Intelligence, Inc., Nashville, TN 37211*
[2] *Fisk University, Nashville, TN 37208*

**Abstract**

In this paper the growth of a 16mm diameter Ce-doped $Tl_2NaYCl_6$ and its characterization as a gamma-ray detector are reported. With a ∅16×8 mm cylindrical sample, energy resolution of 4.1% (FWHM) at 662 keV and light yield of 27,800ph/MeV are measured. Decay times of 91 ns (34%), 462 ns (52%), and 2.1 μs (15%) are calculated. The x-ray excited emission spectrum exhibits bands that are similar to other Tl-based elpasolite scintillators like $Tl_2LiYCl_6$:Ce.

## I. Introduction

Demand for high light yield, high density, and fast scintillators necessitate a continuous search for new materials. Traditional scintillators such as Tl-doped sodium iodide (NaI) and cesium iodide (CsI) have been very reliable standards, supported by decades of research and proven performance. However, various new applications require brighter materials that also have higher densities and faster decay times. For nearly two decades emerging new scintillators such as $LaX_3$ [1, 2], $CeX_3$[3, 4], and $Cs_2AX_5$[5], where A = La or Ce, and X = Cl, Br, or I (halides), have demonstrated the potentials of these metal halides as next-generation scintillation detectors. Rediscovered europium-doped $SrI_2$, with a light yield as high as 110,000 ph/MeV and moderate density of 4.55 g/cm$^3$, has also shown the potential of alkaline metal halide scintillators [6, 7].

Recently Tl-based scintillators have attracted good attention from worldwide scintillator researchers and have given rise to exploration of new scintillation materials in a new direction. These compounds have been investigated and very promising initial results have been published. Tl-based metal halides such as intrinsic scintillators $TlMgCl_3$ (TMC) and $TlCaI_3$ (TCI), as well as Ce-doped $Tl_2LaCl_5$ (TLC) have also been investigated and yielding good results [8, 9, 10]. Published results show that adding Tl to or substituting Tl for other (metal) alkali component in the compounds increases the density and $Z_{eff}$ and improves scintillation performance. For example, in the case of $SrI_2$[6, 7] and $TlSr_2I_5$[11]material density increases from 4.55 to 5.30 g/cm$^3$, $Z_{eff}$ increases from 49 to 61, and the primary decay time of $TlSr_2I_5$ is shorter than $SrI_2$ by about 0.5.$TlSr_2I_5$ with 3% $EuI_2$ doping was recently investigated and reported in [11] and found to have good radiometric properties such as 4% energy resolution at 662 keV, as well as a high light yield of 70,000 ph/MeV [11]. During the same time of the study in [12], we independently investigated and grew intrinsic $TlSrI_3$ and 5% Eu-doped $TlSr_2I_5$ (TSI) [12].

Previously$Cs_2NaYCl_6$:Ce (CNYC),a compound belonging to the elpasolite crystal family, was grown, and reported as a potential fast neutron detector[13].Fast neutron energy spectra were measured under continuous excitation of $^{252}$Cf and $^{241}$Am/Be. Pulse shape discrimination between γ-rays emitted by $^{22}$Na and fasts neutrons emitted by unmoderated $^{241}$Am/Be was also exhibited. Gamma-ray characterization of CNYC was also discussed. An energy resolution of 7% (FWHM) at 662 keV was measured. The light yield was reported to be lower than 10,000 ph/MeV. The material showed a comparable proportionality when compared to $Cs_2LiYCl_6$:Ce (CLYC) and better than NaI:Tl, as expected from elpasolites [13].In this paper the growth of a 16 mm-diameter Ce-doped $Tl_2NaYCl_6$ and its characterization as a gamma-ray detector are reported. Results on energy resolution, relative light yield, non-proportionality behavior and luminescence decay time are reported in the following sections.

## II. EXPERIMENTAL METHODS

<u>*Growth:*</u> Stoichiometric amounts of TlCl, NaCl, and $YCl_3$,with $CeCl_3$ as dopant (all in powder-bead forms), were placed in a clean ∅16mm inner diameter quartz ampoule that was subsequently sealed under high vacuum and placed in a two-zone vertical Bridgman furnace. Furnace zone temperatures were set to a few degrees above melting point at 620°C. Crystal growth commenced at a rate of 2cm/day and post-crystallization cooling at a rate of 250°C/day.

<u>*Sample Preparation:*</u> Samples were retrieved by slicing the boule with a diamond wire saw. To prepare the samples for characterization, the samples were dry-lapped and -polished with SiC sandpapers inside an inert-atmosphere glove box. When sample processing was conducted outside of the glove box, the samples were lapped and polished with mineral oil that acted both as a lubricant as well as a crystal protecting from humidity. As with many other Tl-based elpasolite scintillators, TNYC:Ce is hygroscopic.

<u>*Characterization:*</u> The polished samples were tested for their radiometric and scintillation properties. For x-ray radioluminescence, an x-ray tube source providing <30 keV x-rays was employed, and emission was collected by a fiberoptic-coupled Ocean Optics spectrometer, employing a silicon CCD readout. For measurements of spectroscopic performance, each sample was measured in mineral oil inside a quartz cup that wrapped with Teflon tape as a reflector. Using optical grease, the oil cup was coupled to a R6231-100 Hamamatsu super bi-alkali photomultiplier tube (PMT), coupled to standard nuclear instrument module (NIM) equipment. With this setup spectra from $^{22}$Na, $^{57}$Co, $^{60}$Co, $^{133}$Ba, $^{137}$Cs, and $^{241}$Am were collected, and energy resolution, relative light yield data, as well as non-proportionality behavior were determined. To obtain luminescence decay time information signals from the PMT anode were recorded with a digitizer(CAEN DT5720C) and the waveforms were analyzed offline.

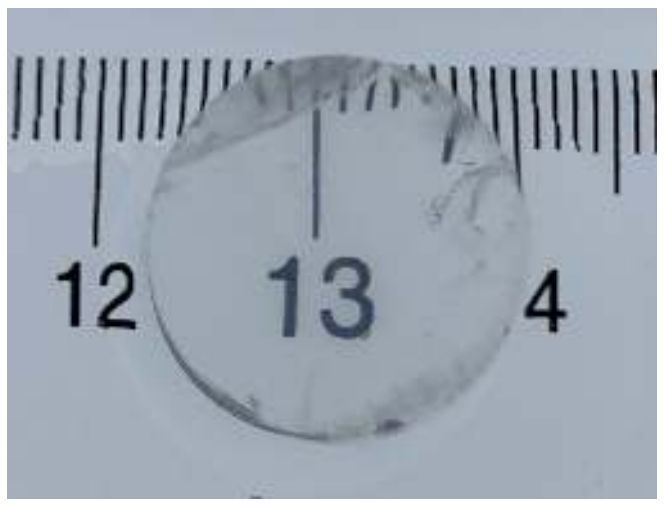

Fig. 1. A single and transparent ∅16mm × 8mm $Tl_2NaYCl_6:Ce^{3+}$ crystal.

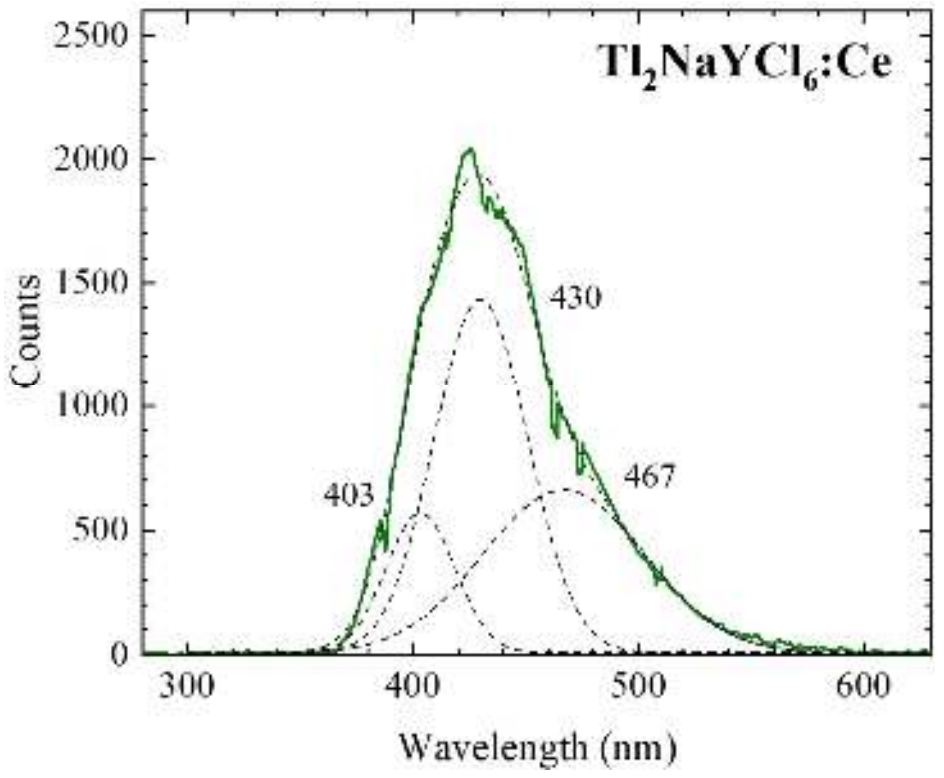

Fig. 2. X-ray excited luminescence spectrum of Ce-doped $Tl_2NaYCl_6$.

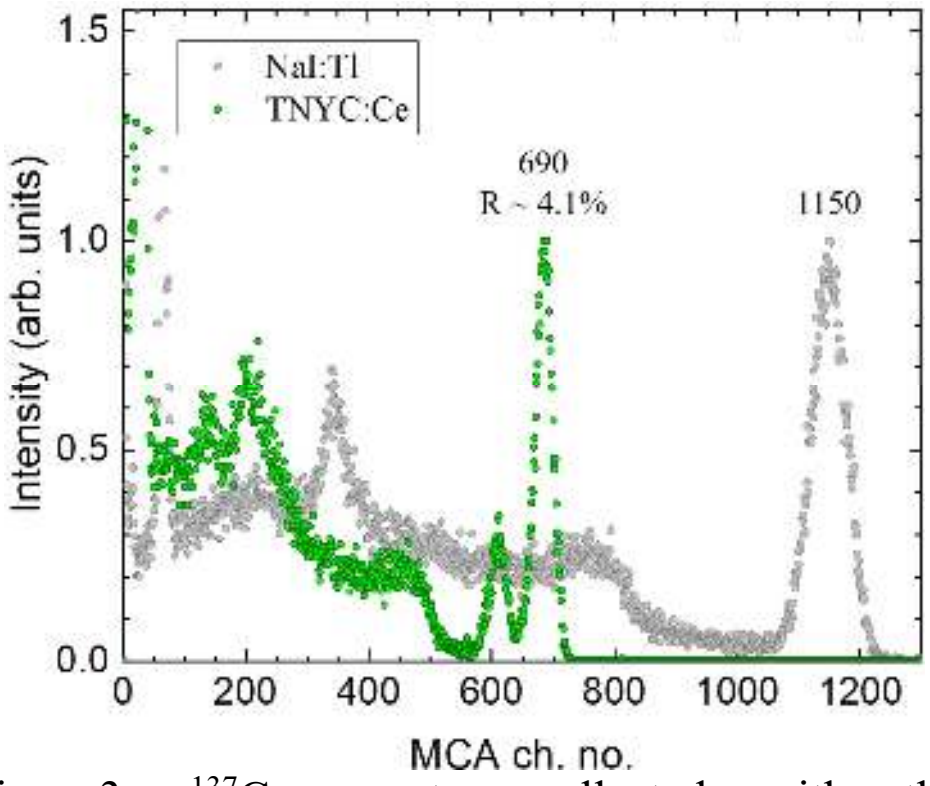

Fig. 3. $^{137}$Cs spectra collected with the $Tl_2NaYCl_6$:Ce sample and with a ∅1″×1″NaI:Tl.

## III. RESULTS AND ANALYSIS

A single and transparent ∅16mm x 8mm of $Tl_2NaYCl_6:Ce^{3+}$ crystal is shown in Fig. 1. Same sample that was characterized for this paper.

### *A. X-ray excited luminescence spectrum*

Luminescence spectrum (Fig. 2) was collected from a sample that was processed in the moments prior to irradiation with 30-keV x-rays. The emission spectrum peaked at about 430 nm similar to the emission spectrum collected from TLYC:Ce [14].

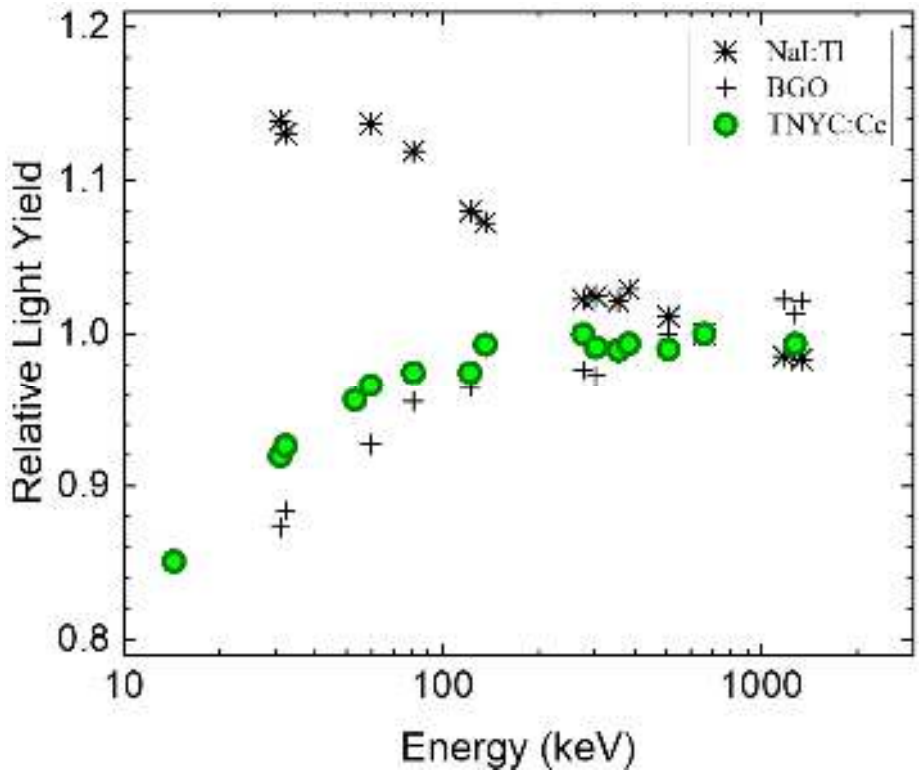

Fig. 4. Non-proportionality data for $Tl_2NaYCl_6$:Ce and compared to NaI:Tl and BGO crystals.

### B. [137]Cs spectrum, energy resolution, and light yield

Fig. 3 shows the spectra of $^{137}$Cs collected by the $Tl_2NaYCl_6$:Ce sample and by a ∅1″×1″NaI:Tl. For $Tl_2NaYCl_6$:Ce the energy resolution of the full energy peak was calculated to be 4.1% (FWHM). From comparison with NaI:Tl (assumed to have 38,000 ph/MeV while considering the quantum efficiency of the R6231-100 PMT), the light yield was calculated to be 27,800 ph/MeV.

### C. Non-proportionality behavior

Spectra of $^{22}$Na, $^{57}$Co, $^{60}$Co, $^{133}$Ba, $^{137}$Cs, and $^{241}$Am were collected with the $Tl_2NaYCl_6$:Ce sample. Peak positions from the full energy peaks and select characteristics x-rays were calculated and used to determine the relative light yield information as a function of energy. Fig. 4 shows the non-proportionality data for $Tl_2NaYCl_6$:Ce and compared to the data for a ∅1″×1″NaI:Tl and 1 cm$^3$ BGO crystals.

### D. Decay times

The average of the waveforms collected at the PMT's anode due to $^{137}$Cs γ-ray excitation of $Tl_2NaYCl_6$:Ce under was analyzed to determine decay timesby fitting three exponential functions to the averaged signals:91 ns (34%), 462 ns (52%), and 2.1 µs (15%).

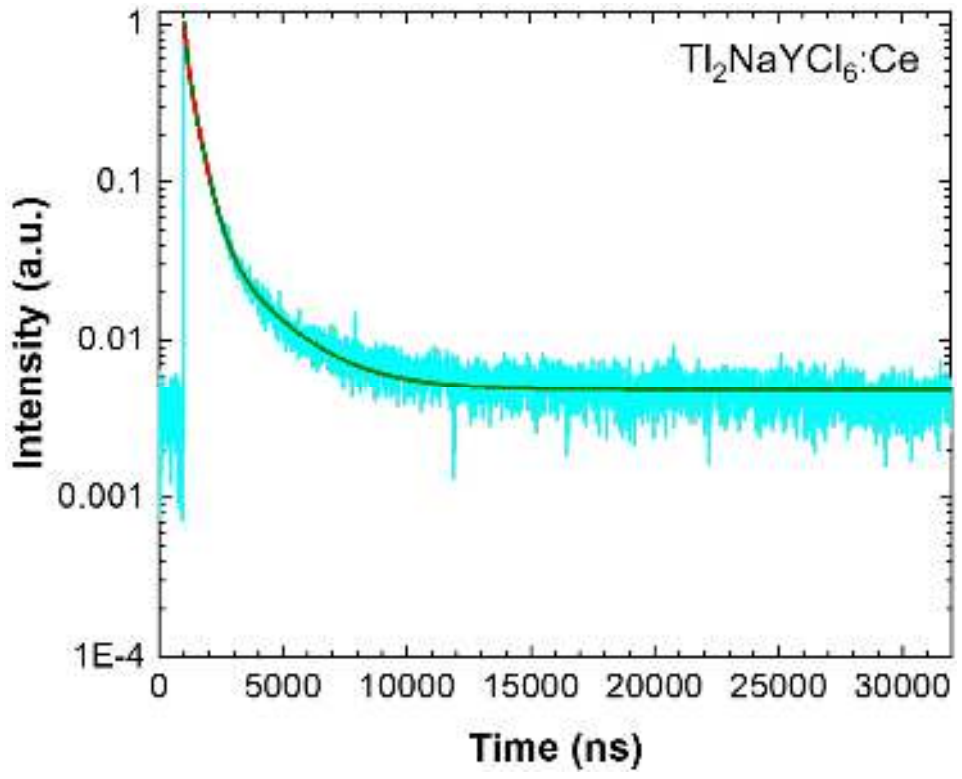

Fig. 5. The averaged waveforms collected at the PMT's anode were analyzed to determine the time profile and the decay times of $Tl_2NaYCl_6$:Ce, which were 91 ns (34%), 462 ns (52%), and 2.1 µs (15%).

## IV. Conclusions

We successfully grew a16-mm diameter Tl-based Ce- doped$Tl_2NaYCl_6$ elpasolite compound. A∅16×8 mm cylindrical sample was extracted, processed, and then coupled to a PMT for characterization as a gamma-ray detector. An energy resolution of 4.1% (FWHM) at 662 keV and light yield of 27,800ph/MeV were measured. Decay times of 91 ns (34%), 462 ns (52%), and 2.1 µs (15%) were calculated.

Table 1 shows the properties of $Tl_2NaYCl_6$:Ce compared to those of other Ce-doped elpasolites$Cs_2NaYCl_6$[13], and $Tl_2LiYCl_6$[14]. The measured properties of $Tl_2NaYCl_6$:Ce were like those of $Tl_2LiYCl_6$:Ce. And based on the results of $Cs_2NaYCl_6$:Ce, $Tl_2NaYCl_6$:Ce is also expected to be a potential fast neutron detector material. Its characterization as a neutron detector will be published in future papers.

Table 1. Properties of $Tl_2NaYCl_6$:Ce, $Cs_2NaYCl_6$:Ce, and $Tl_2LiYCl_6$:Ce.

| **Properties** | **$Tl_2NaYCl_6$:Ce** | **$Cs_2NaYCl_6$:Ce** | **$Tl_2LiYCl_6$:Ce** |
|---|---|---|---|
| Density (g·cm$^3$) | TBD | TBD | 4.5 |
| $Z_{eff}$ | 63 | 41 | 69 |
| Peak Emission (nm, x-ray excited) | 430 | N/A | 430 |
| Energy resolution (% (FHWM), 662 keV) | 4.1 | 7 | 4.2 |
| Light yield (ph/MeV, 662 keV) | 27,800 | > 10,000 | 26,000 |
| Decay times (γ-ray excitation) | 91 ns<br>421 ns<br>2.1 µs | 40 ns<br>530 ns<br>5.5 µs | 55 ns<br>431 ns<br>1.1 µs |
| Neutron detection | Yes | Yes | Yes |

## V. ACKNOWLEDGMENTS

The Authors thank Fisk University's Materials Science and Applications Group for the support.